\documentclass[twocolumn,prb,showpacs]{revtex4}

\usepackage{epsfig,amsmath,amssymb,color}
\begin{document}

\title{Class of variational ansaetze for the ``spin-incoherent'' ground-state of a Luttinger liquid coupled to a spin bath}
\author{Mohammad Soltanieh-ha}
\affiliation{Department of Physics, Northeastern University, Boston, Massachusetts 02115, USA}
\author{Adrian E. Feiguin}
\affiliation{Department of Physics, Northeastern University, Boston, Massachusetts 02115, USA}

\date{\today}
\begin{abstract}
Interacting one-dimensional electron systems are generally referred to as ``Luttinger liquids'', after the effective low-energy theory in which spin and charge behave as separate degrees of freedom with independent energy scales.
The ``spin-incoherent Luttinger liquid'' describes a finite-temperature regime that is realized when the temperature is very small relative to the Fermi energy, but larger than the characteristic spin energy scale. Similar physics can take place in the ground-state, when a Luttinger Liquid is coupled to a spin bath, which effectively introduces a ``spin temperature'' through its entanglement with the spin degree of freedom. We show that the spin-incoherent state can be exactly written as a factorized wave-function, with a spin wave-function that can be described within a valence bond formalism. 
This enables us to calculate exact expressions for the momentum distribution function and the entanglement entropy.
This picture holds not only for two antiferromagnetically coupled $t-J$ chains, but also for the $t-J$-Kondo chain with strongly interacting conduction electrons. We argue that this theory is quite universal and may describe a family of problems that could be dubbed ``spin-incoherent''.
\end{abstract}


\maketitle

\section{Introduction} 

The physics of correlated low dimensional systems is quite different than their higher dimensional counterparts. In three-dimensions, the physics can be described within Fermi liquid theory, that states that there is a one-to-one correspondence
between the excitations of a weakly interacting Fermi system, so-called
quasi-particles, and the excitations of a non-interacting one. Quasi-particles preserve the same quantum numbers as the original excitations in the original system. This scenario breaks down in one dimension (1D): in
this case, the Fermi surface reduces to two points in momentum space, at
$k=\pm k_F$, and the resulting nesting, pervasive at all densities,
prevents the application of perturbation theory. This leads to a
new paradigm: the Luttinger liquid \cite{Haldane1981,Gogolin,GiamarchiBook}. In a Luttinger liquid (LL), the natural
excitations are collective density fluctuations, that carry either spin
(``spinons''), or charge (``holons''). These excitations have different
dispersions, and obviously, do not carry the same quantum numbers as the
original fermions. This leads to the spin-charge separation picture, in
which a fermion injected into the system breaks down into excitations carrying different quantum numbers, each with a characteristic energy scale and velocity (one for the charge, one for the spin).

Recently, a previously overlooked regime at finite temperature has come to
light: the ``spin-incoherent Luttinger liquid'' (SILL) \cite{Matveev2004,Fiete2004,Cheianov2004,Cheianov2005,Fiete2007b,Halperin2007}.
If the spinon bandwidth is much smaller than the holon bandwidth, a small temperature relative to the Fermi energy may actually be felt as a very large temperature by the spins. In fact, the charge will remain very close to the {\it charge} ground-state, but the spins will become totally incoherent, effectively at infinite temperature.
 This regime is characterized by universal properties in the transport, tunneling density of states, and the spectral functions \cite{Fiete2007b}.

In Ref.\onlinecite{Feiguin2009d}, it was shown how this crossover from spin-coherent to spin-incoherent is characterized by a transfer of spectral weight. Remarkably, the photoemission spectrum of the SILL can be understood by assuming that after the spin is thermalized, the charge becomes spinless, with a shift of the Fermi momentum from $k_F$ to $2k_F$. In a follow-up paper \cite{Feiguin2011}, it was shown that a coupling to a spin bath can have a similar effect as temperature, but in the ground-state. The ``spin-incoherent'' ground-state will have the same qualitative features as the SILL at finite temperature. In this work we formalize this conjecture into a unified theory that describes the spin-incoherent ground-state for a variety of model Hamiltonians, such as the $t-J$-Kondo chain and $t-J$ ladders. The main ingredient for the validity of this theory is to have a very flat spinon dispersion, which corresponds to the limit in which spin and charge completely decouple from each other. This formalism is exact in this limit, and provides a new theoretical framework to understand spin-incoherent physics, including the structure of the Kondo lattice ground-state and entanglement.

We start our study by considering an isolated chain of strongly interacting fermions, described by a Hubbard Hamiltonian, or equivalently, by the $t-J$ model in one dimension:
\begin{equation}
H=-t \sum_{i=1,\sigma}^L \left(c^\dagger_{i\sigma} c_{i+1\sigma}+\mathrm{h.c.}\right)
+ J \sum_{i=1}^L (\vec{s}_i \cdot \vec{s}_{i+1} -\frac{1}{4} n_i n_{i+1} ),
\label{H_t-J}
\end{equation}
with the implicit constraint forbidding double-occupancy. Here, $c^\dagger_{i\sigma}$ creates an electron of spin $\sigma$ on the
$i^{\rm th}$ site along a chain of length $L$. The exchange energy is parametrized by $J$, and we take the inter-atomic distance as unity. We express all energies in units of the hopping parameter $t$.

In the $J=0$ limit, the ground-state of this Hamiltonian can be described by the Ogata and Shiba's factorized wave-function \cite{Ogata1990},
which is the product of a fermionic wave-function $|\phi\rangle$, and a spin wave function
$|\chi\rangle$
\begin{equation}
|\mathrm{g.s.}\rangle=|\phi\rangle\otimes |\chi\rangle.
\label{gs}
\end{equation}
The first piece, $|\phi\rangle$, describes the charge degrees of freedom, and is
simply the ground-state of a one-dimensional tight-binding chain of $N$ non-interacting spinless fermions. The spin wave-function $|\chi\rangle$ corresponds to a ``squeezed'' chain of $N$ spins, where all the unoccupied sites have been removed.
In this limit, 
the charge and the spin are governed by independent Hamiltonians.
Since the spin energy scale is determined by $J$, and for the rest of this work we take $J=0$, the spin states are degenerate, and the charge dispersion becomes that of a non-interacting band $\epsilon(k)=-2t\cos(k)$. However, any finite value of $J$ will lift this degeneracy and give the spin excitations a finite bandwidth. Notice that in finite systems, the spin degree of freedom affects the charge through an effective magnetic flux, which in the examples shown here is always identically zero \cite{Caspers1989,Penc1997,Rincon2009}.

Let us assume that we antiferromagnetically couple our chain to a bath of spins. Regardless of the internal structure and dynamics of the bath, it is easy to realize that the charge will be in principle unaffected by it, while the spin degree of freedom will get entangled into a many-body state with the spins from the bath. If we trace over the bath, we expect the spins of the chain to be at an effective finite temperature, parametrized by the magnitude of the system-bath coupling (even though the entire chain plus bath are in a pure state: the ground state of the Hamiltonian). Therefore, the spins of the chain can be driven incoherent by this interaction, while the charge remains in the ground-state. This physics is completely analogous to the SILL physics at finite temperature. We should point out that the coupling with the bath may introduce a gap in the excitation spectrum, but it is to expect that in the regime of interest the gap would remain exponentially small, with the aforementioned picture basically unchanged (whether there is a gap, and/or a critical value of couplings to open a gap is beyond the scope of this work). In Ref.\onlinecite{Feiguin2011} it was numerically shown that this physics is indeed realized in the $t-J$-Kondo chain with strongly correlated conduction electrons, where the Kondo impurities act as an effective spin bath. In this work we calculate the exact ground-state of this system in this limit, and also coupled $t-J$ chains, and we show that the structure of the ground-state is quite universal, and indicates the path toward a unified formalism to describe spin-incoherent behavior at zero and finite temperatures.

\section{Coupled $t-J$ chains}

Let us assume two chains governed by the Hamiltonian (\ref{H_t-J}), and we take the $J \rightarrow 0$. In the limit in which the chains are independent, the exact ground-state will be that of two decoupled factorized wave-functions of the form (\ref{gs}):
\begin{equation}
|\mathrm{g.s.}\rangle=|\mathrm{g.s.}\rangle_1 \otimes |\mathrm{g.s.}\rangle_2=|\phi\rangle_1 \otimes |\phi\rangle_2 \otimes |\chi\rangle_1 \otimes |\chi\rangle_2,
\label{gs2}
\end{equation}
where the subindex $\lambda=1,2$ refers to the chain index.
Now we introduce a small but finite antiferromagnetic interaction between the chains of the form
\begin{equation}
H'=J'\sum_{i=1}^L  \vec{s}_{i,1} \cdot \vec{s}_{i,2},
\end{equation}
where $J'$ parametrizes the interaction perpendicular to the direction of the chains, along the rungs of a ladder. This is equivalent to a $t-J$ ladder without inter-chain hopping. We acknowledge that this is a very idealized scenario, since the presence of exchange always implies the existence of a hopping, since $J' \sim t'^2$. Still, this could be considered a model for a two band problem with a Hund coupling, as studied in Ref.\cite{Tsvelik}. It is to expect that this interaction will couple the spin pieces of the wave-function $|\chi\rangle_1$ and $|\chi\rangle_2$, leaving the charge unaltered:
\begin{equation}
|\mathrm{g.s.}\rangle=|\phi\rangle_1 \otimes |\phi\rangle_2 \otimes |S\rangle,
\label{gs3}
\end{equation}
where $|S\rangle$ represents the many body state of the spins for the two coupled chains, once they become entangled by action of the Hamiltonian $H'$.
It is also to expect that this state $|S\rangle$ will be a singlet. However, the exact structure of this singlet is not necessarily trivial. 

\begin{centering}
\begin{figure}
\epsfig{file=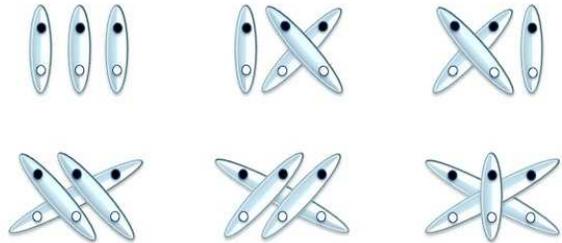,width=80mm}
\caption{Possible singlet coverings for two sublattices, and 3 spins per sublattice.} 
\label{fig:VB1}
\end{figure}
\end{centering}

Without attempting to deduce the exact effective Hamiltonian for the spin sector, we shall propose a variational ansatz for the wave-function, that we later prove to be exact by numerical means. We argue that every time two spins interact on a rung via $H'$, they will become entangled forming a singlet. Since the interaction along the chain is set to $J=0$, these spins will remain entangled as they move apart from each other by action of the hopping term. Therefore, this entanglement persists at infinite distance. If we consider a system with periodic boundary conditions, it is to expect that eventually all spins from one chain will interact with all the spins on the second chain. In a one dimensional system, they cannot hop past each other, but in a chain with periodic boundary conditions, they can wind around the boundaries and come from the other side. Therefore, we will have a superposition of singlets that connect all possible pairs of spins on both chains, and at all distances, with the same amplitude. In order to describe this wave-function it is useful to resort to a valence bond(VB) picture \cite{Pauling1933,Oguchi1989,Beach2006,Tang2011}. Let us assume that each chain corresponds to a sublattice. Then, our wave-function is the equal superposition of all possible valence-bond coverings connecting the two sublattices, as shown in Figure \ref{fig:VB1} for the particular case of three electrons per chain.

\begin{centering}
\begin{figure}
\epsfig{file=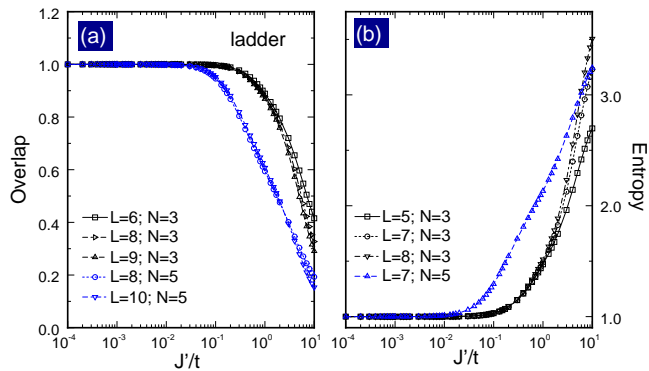,width=50mm,angle=-90}
\caption{Exact diagonalization results for small ladders of length $L$, and $N$ electrons per chain: (a) overlap with variational wave-function, and (b) entanglement entropy between chains, normalized by the exact value for $J'\rightarrow0$.} 
\label{fig:ladder}
\end{figure}
\end{centering}

In order to prove that our guess accurately describes the physics, we have numerically computed the overlap between the exact ground-state and the variational ansatz on small systems with periodic boundary conditions. We assumed $S^z_{Total}=0$ and taken the number of particles not a multiple of $2$, to avoid degeneracies. We show the results in Figure \ref{fig:ladder}(a), for different values of $J'$. The overlap is $1$ within numerical precision for a range of small values of $J'$. As $J'$ increases, we observe how this overlap becomes smaller, but still remains higher than $0.9$ for $J' < 0.1$. This range depends only on the number of conduction electrons $N$, and tends to get smaller as the $N$ increases. 

Having shown that the ansatz is a good description of the spin-incoherent regime for $J' \rightarrow 0$, we proceed to derive some straight-forward exact results that can be obtained using the variational form of the wave function. For a start, the entanglement between chains originates from the spin, and the charge does not contribute to it. One might feel inclined to think that spins are in a maximally entangled state. However, we should not forget that the VB basis is overcomplete, and in fact, the entanglement entropy is not $S=N\log2$, as one might expect for a state with $N$ singlets. Using the exact wave-function it is relatively easy to obtain a closed expression for $S$, as shown in the Appendix: 
\[
S=\log(N+1).
\]
Looking at this expression more closely, one realizes that this is equivalent to two spins $S=N/2$ in a maximally entangled state, instead of $N$ spins $1/2$ in a maximally entangled state (see Appendix). 
In Figure \ref{fig:ladder}(b) we show the entanglement entropy, normalized by the exact value for $J'=0$. Same as the overlap, the expression holds for a range of $J'$, and $S$ increases as the charge becomes also entangled with the spin. 

\begin{centering}
\begin{figure}
\epsfig{file=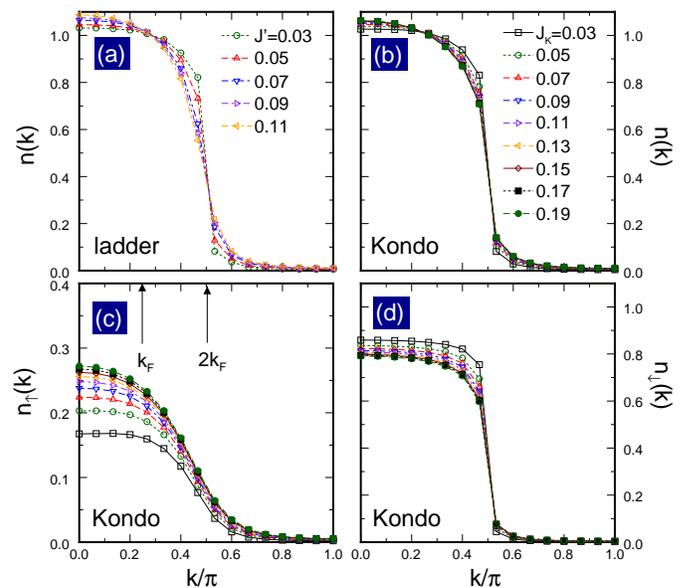,width=80mm,angle=-90}
\caption{Momentum distribution function for (a) coupled $t-J$ chains as a function of the inter-chain coupling $J'$, and (b),(c),(d) the Kondo lattice, as a function of the Kondo coupling $J_K$. The two lower panels show the results for different spin orientations. Calculations were done with DMRG for a system of size $L=30$ and $N=15$ conduction electrons, and periodic boundary conditions.} 
\label{fig:mdf}
\end{figure}
\end{centering}

It is enlightening to calculate the momentum distribution function (MDF) for the fermions:
\begin{equation}
n(k) = (1/L)\sum_{l,\sigma} \mbox{exp}(ikl)\langle c^\dagger_{1,\sigma}c_{l,\sigma}\rangle
\label{nk}
\end{equation}
In order to estimate this quantity, we follow Ref.\onlinecite{Penc1997} and break the fermionic operators $c^\dagger_{i,\sigma}$ and $c_{i,\sigma}$ into a spinless fermionic operators $f^\dagger_i$,$f_i$ acting on the (spinless) charge part of the wave-function, and new operators $Z^\dagger_{i,\sigma}$ and $Z_{i,\sigma}$ acting on the spin part of the wave-function. These spin operators have a very peculiar behavior: $Z^\dagger_{i,\sigma}$ inserts a spin $\sigma$ to the spin chain after skipping the first $i-1$ spins and makes it $N+1$ sites long, while $Z_{i,\sigma}$ has the opposite effect, shortening the chain. For instance, for the first site of the chain, we have:
\begin{equation}
c^\dagger_{1,\sigma}=Z^\dagger_{1\sigma}f^\dagger_1.
\end{equation}
The generic expression for the operators can become more complicated, since to act with the $Z$ operators on the spin chain, we need to count the number of charges on the spinless fermion chain. We refer the reader to Refs.\onlinecite{Sorella1991,Pruschke1991,Penc1997,Penc1997b} for details. The action of the operators $c^\dagger_{1,\sigma}c_{l,\sigma}$ is to move a fermion from site $l$ to site $1$. If there are no particles in between, the spin wave-function will remain unchanged. If there is one or more particles in between, it is quite easy to realize that since the spin wave-function is the equal sum of all singlet coverings, it will also remain so after moving one of the ends of a singlet across any number of sites. Therefore, the momentum distribution function reduces to 
\begin{equation}
n(k) = (1/L)\sum_{l,\sigma} \mbox{exp}(ikl)\langle f^\dagger_{1}f_{l}\rangle,
\label{nkb}
\end{equation}
which is nothing else but the MDF for spinless fermions. Since the charge wave-function is that of non-interacting particles, we find that the excitations are free spinless fermions with quasi-particle weight $z=1$, and Fermi momentum $2k_F$. 
In Figure \ref{fig:mdf}(a) we show the MDF calculated for large systems using the density matrix renormalization group (DMRG) method \cite{White1992,White1993}, indicating that the quasi-particle weight may remain finite for a range of $J'$. We have to concede that since the calculations are on finite-systems with $L=30$ sites and $N=15$ electrons per chain, we cannot argue with full certainty that the discontinuity at $k=2k_F$ is not actually a singularity, and this remains an interesting problem to pursue. 

\begin{centering}
\begin{figure}
\epsfig{file=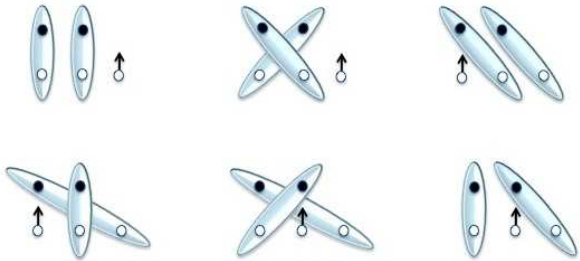,width=80mm}
\caption{Possible singlet coverings for two sublattices with unequal number of sites, and an excess up-spin.} 
\label{fig:VB2}
\end{figure}
\end{centering}

\section{The $t-J$ Kondo chain}

We consider a Kondo chain in which the conduction electrons strongly interact, and are described by Hamiltonian (\ref{H_t-J}). At the same time, the electrons are antiferromagnetically coupled to localized impurities via an exchange $J_K$:
\begin{equation}
H_K=J_K \sum_{i=1}^L \vec{s}_i \cdot \vec{S}_{i},
\label{fullH}
\end{equation}
where  $\vec{s}_i$ describes the conduction spins  and $\vec{S}_i$ the localized spins. It is easy to see that this is equivalent to the two coupled chains, in which one of them is at half-filling. Curiously, this model has not received much attention in the literature. It has been shown that in the limit of $J\rightarrow0$, any infinitesimal $J_K$ will yield a ferromagnetic ground-state\cite{Yanagisawa1994}, in which the localized impurities are underscreened: the $N$ conduction spins will screen $N$ impurity spins, and the remaining ``unpaired'' impurities will be in a ferromagnetic state with maximum spin $S_{Total}=(L-N)/2$. Notice that this means that the paramagnetic state with a large Fermi surface is totally suppressed in this regime. 

\begin{centering}
\begin{figure}
\epsfig{file=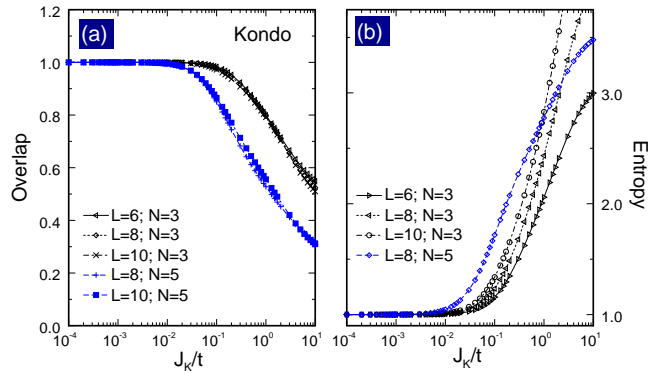,width=50mm,angle=-90}
\caption{Exact diagonalization results for small $t-J$ Kondo chains of length $L$, and $N$ conduction electrons: (a) overlap with variational wave-function, and (b) entanglement entropy between chains, normalized by the exact value for $J_K\rightarrow0$.} 
\label{fig:kondo}
\end{figure}
\end{centering}

This state will be a multiplet, and for convenience we focus on the configuration with projection $S^z_{Total}=S_{Total}$. Following a similar reasoning as in the previous section, we argue that the $N$ conduction electrons will form the same VB state as the one described before, while the unpaired impurities will all point in the same direction. The polarized spins can sit on any site of the lattice with equal probability. Therefore, our ansatz can be written as:
\begin{equation}
|\mathrm{g.s.}\rangle=|\phi\rangle \otimes |S\rangle \otimes |\sigma\rangle,
\label{gs4}
\end{equation}
where $|S\rangle$ is the VB wave function, and $|\sigma\rangle$ indicates the positions of the unpaired polarized spins:
\[
|\sigma\rangle = \sum_{x} |x\rangle,
\]
This wave-function is the sum with equal amplitude of all the configurations $|x\rangle$ of $L-N$ particles in $L$ sites.
Figure \ref{fig:kondo}(a) shows the overlap between the exact and variational ground-states, and we again observe identical behavior as the $t-J$ ladder. 
The VB basis for this problem is overcomplete, and we also have to account for the unpaired polarized spins, as shown schematically in Figure \ref{fig:VB2}. This generalization can be easily carried out\cite{Damle2010}, and it is still straightforward to obtain a closed expression for the entropy, which is slightly more complicated than the one for the coupled chains. We show results in Figure \ref{fig:kondo}(b), which have strong resemblance with those for the ladder. 

The calculation of the MDF is strictly the same as before and the results are identical for $J_K=J'=0$. Notice however, that unlike the $t-J$ ladder, the MDF for up and down spins will be different, and only the sum of the two will be the same. This is shown in Figure \ref{fig:mdf}(b),(c), and (d). In particular, there is a striking difference between the MDF for the majority up and minority down electrons. The up electrons present a clear discontinuity at the Fermi level, while the down electrons display the behavior of a Luttinger liquid with zero quasi-particle weight. The sum of the two, shown in Figure \ref{fig:mdf}(b) of course hides these interesting features. This resembles the behavior of a Fulde-Ferrell-Larkin-Ovchinnikov (FFLO) polarized paired state in one dimension\cite{Yang2001,Orso2007,Feiguin2007c,Feiguin2009b,Heidrich-Meisner2010,Lutchyn2011,Dalmonte2012}. However, the physics of our problem is quite different, since only the spin entangles, and not the charge. We would rather call this state a ``half-Fermi-liquid'', or ``half-Luttinger-liquid''. Whether these are finite-size artifacts or not, undoubtedly, it is a problem that requires further study.
We point out again that the shift to a larger momentum in the MDF should not be confused with a large Fermi surface, that is a feature of the {\it paramagnetic} phase of the Kondo lattice.

\section{Summary and Conclusions}

We have presented a ground-state ansatz for antiferromagnetically coupled $t-J$ chains, and the $t-J$-Kondo chain that is numerically exact in the limit of $J=0$ (corresponding to infinite-$U$ Hubbard chains), and coupling to the bath $J',J_K \rightarrow 0$, as tested on small systems. Moreover, our DMRG results indicate that the variational wave functions describe the physics of the problem in a range of $J'$ and $J_K$. In this regime, the charge and the spin can be considered to a good extent as separate degrees of freedom with independent dynamics: the charge can be described as non-interacting spinless fermions in the ground-state, while the spin is entangled into a VB-like state where all valence bond coverings have the same weight. The inter-chain coupling $J'$ and the Kondo interaction $J_K$ parametrize an effective {\it spin} temperature. If we trace over the bath, the spins of an isolated chain will be in equilibrium at a certain ``quasi-spin temperature''. This spin temperature is not infinite, since we have proven that the spin is not maximally entangled. However, this state seems to correspond to a fine-tuned point in which excitations can be described as free spinless fermions. The momentum distribution functions show a discontinuity at $2k_F$ indicating that the system is no longer a Luttinger liquid. However, it is noteworthy to point out the peculiar behavior of the $t-J$-Kondo chain. Since it has a polarized ferromagnetic ground-state, the up and down fermions behave notoriously different: at finite $J_K$, the minority spins show a Luttinger liquid-like behavior, while the majority fermions appear as almost free particles. 
 Notice that the ground-state of the $t-J$-Kondo chain is a spin multiplet, and this ``half-Luttinger-liquid'' behavior in the $t-J$-Kondo chain may be an artifact of working on the maximally polarized state. However, we believe that this physics deserves further investigation.

In the ferromagnetic phase of the conventional Kondo chain, the natural excitations are spin-polarons formed by a conduction electron and impurities forming a bound state that propagates coherently \cite{Ueda1991,Smerat2009,Smerat2011}. In our scenario, charges and impurities cannot form a bound state because the charge decouples completely from the spin. However, the spin of the electrons remains entangled with the impurities at all distances, but this does not imply an effective potential acting on the charges that can still move freely.

An interesting observation is that even though the Hamiltonians are local and the charge degree of freedom is totally uncorrelated, the spins remain correlated at {\it infinite distance}, and the spin-spin correlations between chains are constant at all distances (see Appendix). 

The introduced wave-functions establish a framework to study spin-incoherent behavior in systems with spin-charge separation. Normally considered a finite-temperature scenario, this physics can also be realized at zero temperature, once the system is coupled to external spin degrees of freedom. It is not restricted to the models used in this work for illustration, but the theory can be easily extended and generalized to other cases, such as an arbitrary number of coupled $t-J$ chains, for instance \cite{Anderson1990,Putikka1994}. 

We point out that even though our study applies to systems with periodic boundary conditions, same ideas apply to problems with open boundary conditions. In that case, we expect the VB wave-function to be quite different, with only the first kind of configurations shown in Figure \ref{fig:VB1} carrying most of the weight, and a consequent entanglement entropy $S=N\log{2}$ corresponding to infinite effective spin temperature \cite{Feiguin2011}. 

Contrary to other problems studied with VB-type variational wave-functions \cite{Sandvik2012}, the accuracy of our ansaetze seems to depend primarily on the number of particles $N$, and not the size of the chains $L$, which may suggest that our description will still be valid for large systems, as long as the density is sufficiently small. Yet, our DMRG results a quarter-filling still display the same physics. In any case, one has to keep in mind that our considerations are strictly valid in the limit $J',J_K\rightarrow 0$.


 An important issue that we have not addressed in this paper is the nature of the actual spin Hamiltonian governing the dynamics of the spins. This is an interesting problem and for the moment it remains open.

\section*{Acknowledgments}
We thank A. Aligia, C. Batista, F. Essler, G. Fiete, A. Sandvik, and M. Troyer for stimulating discussions and useful comments.
AEF is grateful to NSF for funding under grant DMR-0955707.



\section*{APPENDIX: ENTANGLEMENT ENTROPY AND CORRELATIONS}

In this section we calculate the entanglement entropy between the conduction electrons in one chain, and the bath. In the $t-J$ ladder, the bath is modeled by a second chain, while in the $t-J$ Kondo model, by localized impurities. As illustrated in Figures \ref{fig:VB1} and 
\ref{fig:VB2}, we squeeze the chain by removing all the unoccupied sites, reducing the configuration space to a spin problem with no charge. In these cartoons, the black dots and white dots will be referred to as A and B sub-lattices. Each sublattice will have $N$ sites, instead of $L$, where $N$ is the number of electrons in the chain. 

We start by considering the $t-J$ ladder, in which both sublattices have the same number of sites, and the ground-state is represented by all possible VB coverings connecting the two.  
Instead of using the overcomplete  VB basis for the calculation, we will work in the space of spin configurations. In this basis, the states can be classified by the number $N_\downarrow$ of down spins in sublattice A. Since the total spins projection $S^z$ is conserved, this also fixes the number of down spins on sublattice B. The coefficient in front of each configuration is then given by
\begin{equation}
g(N,N_\downarrow)=N_{\downarrow} !(N-N_{\downarrow})! \times (-1)^{N_{\downarrow}},
\label{deg}
\end{equation}
which counts the number of times each of them is repeated in the ground-state, times a sign arising from the singlets (we have ignored the normalization for the time being).

The Von Neumann entanglement entropy $S$ is defined as
\begin{equation}
S_\mathrm{A}=-\mathrm{Tr}\left( \rho_\mathrm{A}\log{\rho_\mathrm{A}} \right),
\label{S_A}
\end{equation}
where $\rho_\mathrm{A}$ is the reduced density matrix for sublattice A, obtained by tracing over the states on sublattice B.
It is easy to see that $\rho_\mathrm{A}$ can be separated into blocks, each labeled by $N_\downarrow$. Since $N_\downarrow$ can assume values $N_\downarrow=0,\cdots,N$, the number of such blocks is $N+1$. The linear dimension for each block is given by the number of possible arrangements of $N_\downarrow$ spins in $N$ sites:
\[
d(N,N_\downarrow)=\frac{N!}{N_\downarrow ! (N-N_\downarrow)!}.
\]

It is easy to see that, since all configurations with fixed $N_\downarrow$ will appear with the same coefficient, each block will have all matrix elements equal to $\rho_\mathrm{A}(N,N_\downarrow)_{i,j}=d(N,N_\downarrow) g^2(N,N_\downarrow)$:
\begin{equation}
\rho_\mathrm{A}(N,N_\downarrow)=d(N,N_\downarrow) g^2(N,N_\downarrow)
\left(
\begin{array}{cccc}
1 & 1 & \cdots & 1 \\
1 & 1 & \cdots & 1 \\
\vdots & \vdots & \ddots & \vdots \\
1 & 1 & \cdots & 1
\end{array}
\right)
\end{equation}
This matrix has only a single non-zero eigenvalue $w(N,N_\downarrow)=d^2(N,N_\downarrow)g^2(N,N_\downarrow)=(N!)^2$, the same for all blocks.
Finally, the full matrix has to be normalized such that $\mathrm{Tr}(\rho_\mathrm{A})=1$. Therefore, we obtain $N+1$ blocks, each with a single non-zero eigenvalue $w=1/(N+1)$. Hence, the entanglement entropy (\ref{S_A}) is given by:
\[
S_\mathrm{A}=\log{(N+1)},
\]
which is our final result. 

This expression is equivalent to two spins $S=N/2$ in a maximally entangled state, where each spin is obtained by the addition of the $N$ spins $1/2$ of each sublattice. 
This analogy can be made rigorous by observing that the spin wave-function is the ground-state of the Hamiltonian:
\[
H_\mathrm{AB}=\sum_{i,j} \vec{s}_{i,\mathrm{A}}\cdot \vec{s}_{j,\mathrm{B}} = \vec{S}_\mathrm{A} \cdot \vec{S}_\mathrm{B},
\]
with $\vec{S}_\mathrm{A} = \sum_i \vec{s}_{i,\mathrm{A}}$, and a similar expression for sublattice B. The ground state is a singlet of two spins $S=N/2$, a maximally entangled state. Notice that this is not the actual spin Hamiltonian for the coupled $t-J$ chains, since the spectra are different. Now we can make use of this solution to calculate the spin-spin correlations. The Hamiltonian can be re-written as:
\[
H_\mathrm{AB}=\frac{1}{2}\left[(\vec{S}_\mathrm{A}+\vec{S}_\mathrm{B})^2-\vec{S}_\mathrm{A}^2-\vec{S}_\mathrm{B}^2 \right].
\]
From this expression, we obtain the ground-state energy:
\[
\langle H_\mathrm{AB}\rangle = \sum_{i,j} \langle \vec{s}_{i,\mathrm{A}}\cdot \vec{s}_{j,\mathrm{B}}\rangle = -\frac{N}{2}\left(\frac{N}{2}+1\right).
\]
Since all the correlators should be equal, we find:
\[
\langle \vec{s}_{i,\mathrm{A}}\cdot \vec{s}_{j,\mathrm{B}}\rangle = \frac{1}{N^2} \langle H_\mathrm{AB}\rangle = -\frac{1}{4}-\frac{1}{2N},
\]
In order to calculate the correlations in the actual $t-J$ ladder we need to include the charge contribution:
\[
\langle \vec{s}_{i,1} \cdot \vec{s}_{j,2} \rangle = (-\frac{1}{4}-\frac{1}{2N})\langle n_{i,1} n_{j,2} \rangle = (-\frac{1}{4}-\frac{1}{2N})\left(\frac{N}{L}\right)^2,
\]
which indicates that the correlations saturate in the thermodynamic limit.

The calculation for the $t-J$-Kondo lattice follows identical steps, except that since the B sublattice has $L$ sites, the degeneracy for each sector acquires a slightly more elaborate form:
\[
g(L,N,N_\downarrow)=\dfrac{(N-N_{\downarrow})! (L-N+N_{\downarrow})!}{(L-N)!} \times (-1)^{N_{\downarrow}}
\]

In this case, the single non-zero eigenvalues for each sector are given by:
\[ 
w(L,N,N_\downarrow)=d(L,N,N_\downarrow)g^2(L,N,N_\downarrow)d_\mathrm{B}(L,N,N_\downarrow),
\]
where
\[
d_\mathrm{B}(L,N,N_\downarrow)=\frac{L!}{(L-N+N_\downarrow)!(N-N_\downarrow)!}
\]
is the number of configurations in the B sublattice, for each configuration of the A sublattice. Since the eigenvalues are different for each sector, the normalization and the entropy are obtained by adding numerically over the $N+1$ blocks.


\end{document}